\documentclass{article}
\usepackage[utf8]{inputenc}
\usepackage{amsmath}
\usepackage{amssymb}
\usepackage{anysize}
\usepackage{bold-extra}
\usepackage{color}
\usepackage{enumerate}
\usepackage{fancyhdr}
\usepackage{graphicx}
\usepackage[linktocpage,colorlinks,urlcolor=blue]{hyperref}
\usepackage{multirow}
\usepackage[final]{pdfpages}
\usepackage{rotating}
\usepackage{subfig}
\usepackage{textcomp}
\usepackage{url}
\usepackage{verbatim}
\usepackage{wrapfig}
\usepackage{amsfonts}
\usepackage{comment}
\usepackage{epigraph}
\usepackage{slashed}
\usepackage{cancel}
\usepackage{epsf, array, color}
\usepackage{authblk}
\usepackage[
backend=biber,
style=alphabetic,
sorting=ynt
]{biblatex}
\title{Muography applied to archaeology}

\author[1]{Theodoros Avgitas}
\author[2]{Sabine Elles}
\author[3]{Corinne Goy}
\author[2]{Yannis Karyotakis}
\author[1]{Jacques Marteau}
\affil[1]{Institut de Physique des 2 Infinis de Lyon (IP2I), IN2P3, CNRS, Université Lyon 1, UMR 5822, Lyon; France}
\affil[2]{Laboratoire d'Annecy de Physique des Particules (LAPP), Univ. Savoie Mont Blanc, CNRS/IN2P3, Annecy; France.}
\affil[3]{Laboratoire de Physique Subatomique et Cosmologie (LPSC), Université Grenoble Alpes, CNRS/IN2P3, Grenoble INP, Grenoble; France}
\date{ February 2022 \\ Proceedings of the RST  2021 conference}

\begin{document}

\maketitle

\section{Introduction -- Tumuli}
Tumuli were used in the ancient times to cover a funeral monument. Usually people were initially building the monument, whose size and decoration was testifying the importance of the buried people. Then the monument was covered by soil, transferred by the nearby, forming a tumulus. Tumuli, of all sizes, can be found in many places around earth and especially in Northern Greece. It is important for archaeologists to be aware of the existence of an underground funeral monument, before any excavation. Combining muon imaging or muography, with other techniques, can give an answer to this question without modifying and disturbing the site.   

In this paper, Section \ref{sec:muography} introduces the general principles and issues of muography. Section \ref{sec:mc} shortly describes simulation techniques used in high energy physics. Section \ref{sec:data} presents the experimental aspects of a muography application from the detector setup to the reconstruction of muon distributions. Section \ref{sec:results} compares data with the simulation before concluding and presenting some ideas for future developments. 

\section{Muography \label{sec:muography}}

Muon imaging or muography has emerged as a powerful non-invasive method  to complement standard tools in Earth Sciences and is nowadays applied to a growing number of fields such as industrial controls, homeland security, civil engineering. This technique relies on the detection of modifications - absorption or scattering - in the atmospheric muon flux when these particles cross a target. 

Atmospheric muons are secondary products of primary cosmic-rays, essentially protons and helium nuclei expelled by stars, interacting with nuclei encountered on the top of the atmosphere. In a simplified model, the particles cascades are~: primary cosmic-rays + oxygen/nitrogen nuclei → parent mesons (pions/kaons) → secondary muons. 

\begin{figure}
    \includegraphics[width=15cm]{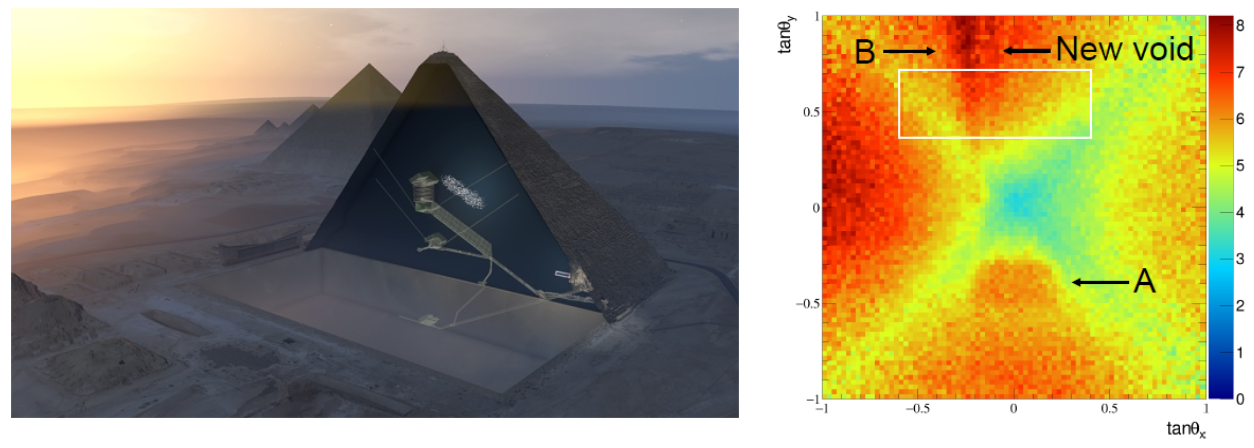}
    \caption{Discovery a new void in the Khufu's pyramid by the ScanPyramids collaboration \cite{Scan}. \textbf{Left:} 3D reconstruction of the void by multiple muons imaging ouside and inside the pyramid. \textbf{Right:} typical density map scatter plot -- muon radiography -- showing the large contrasts between different areas of the pyramid.
    \label{fig:Pyramids}}
    \end{figure}

The rather low interaction cross-section of muons with matter ensures that most of them reach the Earth’s ground level and that furthermore they may significantly penetrate large and dense structures. As suggested originally by Alvarez in 1970 \cite{Alvarez} for the Chephren pyramid, this property may be exploited to perform density contrasts analysis of the interior of the target like X-rays do in medical imaging. Figure \ref{fig:Pyramids} shows the discovery by the ScanPyramids collaboration \cite{Scan} of a new void in the Khufu's pyramid by multiple muon detectors outside and inside the pyramid. 

The range of applications of muography is very large since it may concern any large and dense structures, from volcanic domes to mountains, anthropic buildings etc which remain opaque to the standard investigation methods. 

\subsection{Underlying physics governing muography techniques}
Being charged leptonic particles, atmospheric muons undergo Coulomb electric interactions with the electrons and nuclei they cross along their journey inside matter. This results in a loss of a fraction of their energy by ionisation and radiation and also in a deviation of their trajectory. These properties, sensitive to the density and composition of the target are exploited in the two different modes of muography called “absorption muography” and “scattering muography”. 
The absorption mode is the same as for the X-ray medical imaging. One infers the mass distributions inside a given target from the measurement of the reduced muons flux due to their interaction with the matter of the target. The scattering mode allows the reconstruction of the mass distributions from the measurement of the muons trajectory deviation angles upstream and downstream the target. It is usually restricted to small targets while the absorption mode is well-suited for large volumes imaging.

\subsection{Detection of atmospheric muons} 

\begin{figure}
    \centering
    \includegraphics[width=15cm]{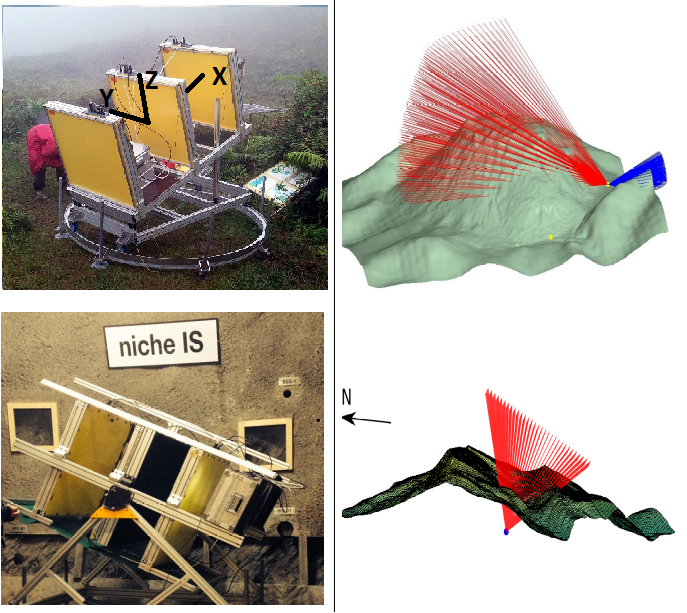}
    \caption{Examples of muography applications. \textbf{Upper row:} open-air installation on the slope of the Soufrière of Guadeloupe (Lesser Antilles). \textbf{Lower row:} underground measurement in the Mont-Terri laboratory (Switzerland). The muon trackers are shown on the left while the muon trajectories falling into their acceptance are illustrated on the sketch on the right.}
    \label{fig:DirectProblem}
    \end{figure}

Detecting muons at the ground level exploits the same properties : charges resulting from the ionisation due to the muons when crossing plastic scintillators, silicium cells or gaseous systems of a detector are collected and shaped into an electric pulse. By recording this signal at several points along the trajectory, the direction of the muon track can be reconstructed: the muography detectors belong to the “trackers” category. 
Figure \ref{fig:DirectProblem} illustrates two experimental implementations where the muon detector (pictures of the left column) is located either on the slope of an active dome (the Soufrière of Guadeloupe, Lesser Antilles, France) or in a gallery of the underground Mont-Terri laboratory (Jura, Switzerland). These trackers use plastic scintillators as detection medium. The sketches in the right column represent all muons trajectories falling into the detectors acceptance.

Tracking performance is measured in terms of spatial and angular resolution, usually driven by the size of the detector segmentation, and in terms of timing resolution. For a large structure muography, an important parameter is the detector acceptance \cite{Sullivan}, i.e. its capability of collecting the maximal number of muons for a given active surface. 
The background rejection is important for outdoor applications where one needs to eliminate, on one hand, random coincidences and requires fine timestamps of the order of the nanosecond or below, and on the other hand the small fraction of atmospheric electrons surviving at the ground level. 

\subsection{From data distributions to density maps }
The most difficult step in muography is the so-called “inverse problem”, ie going from raw data to reconstructed mass distributions 
In absorption mode the detector measures the attenuation of the muon flux integrated all over the path of the muons $\mathcal{L}$ inside the studied target of density $\rho$ , i.e. its “opacity”  defined as $ \varrho=\int_{\mathcal{L}} \rho dl.$
Going from an opacity map to a density map requires therefore a model or more generally an “inversion technique” that provides the most probable mass distribution functions inside the target.
The inverse problem needs to be constrained by the available “a priori” information  but is also driven by the data quality which imposes strict requirements on the detector performance in terms of acceptance, resolution, stability in operation, duty cycle etc. 

There are intrinsic limitations to the muographic inverse problem, the major one being the limited statistics of the measurement : open sky muon flux is of the order of 1 muon per square centimeter per minute and may be reduced by several orders of magnitude after a large target. On top of the statistical limitation, there are intrinsic ambiguities for a single-point measurement since a muon deficit (negative anomaly) or a muon excess (positive anomaly) w.r.t. a given model leads to an infinite number of possibilities as to the precise location of this anomaly along the path. 

The approach reported in this paper to answer to the above problem is based on techniques developed in the field of high-energy physics; these are presented in the following section.

\section{Monte-Carlo methods \label{sec:mc}}

Monte-Carlo technique is the back-bone of the simulation program GEANT4 \cite{GEANT1} widely used in high-energy physics. GEANT4 is able to simulate the interactions of particles, including muons, with matter in any volumes. Monte-Carlo technique relies on the mathematical concept to approach numerical values by using random processes. This approach is particularly well-suited to describe interactions of particles as these interactions are governed by probability density functions.  

\begin{figure}
    \centering
    \includegraphics[width=12.cm]{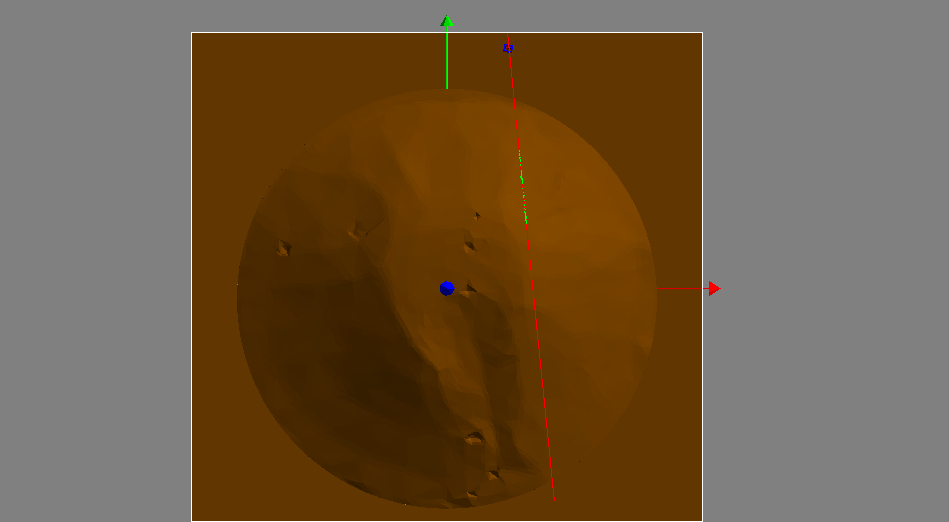}
    \includegraphics[width=12.cm]{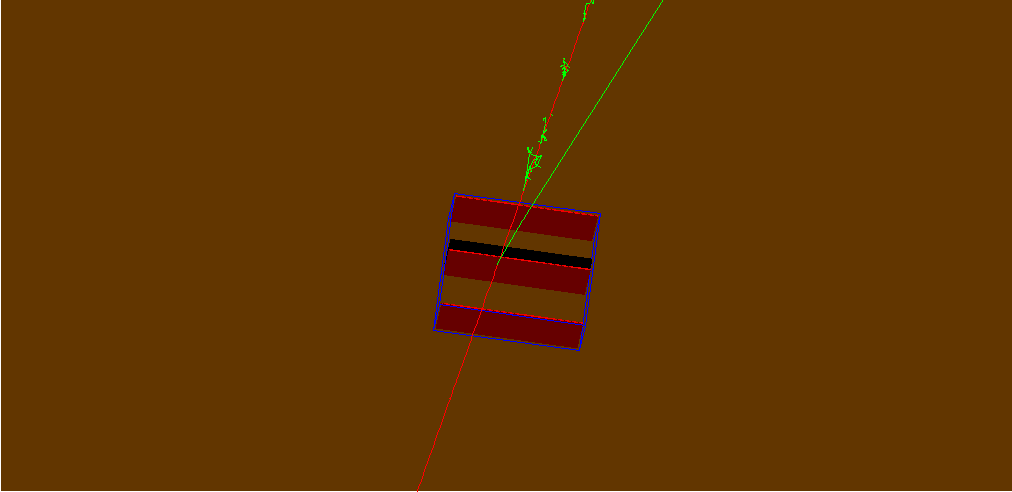}
    \caption{\textbf{Top:} Display of a simulated muon crossing the tumulus. The tumulus is seen from the top: the incident muon represented by the dashed red line crosses the tumulus; the interaction with the tumulus matter generating secondary particles, is materialized by the green dots along the line. \textbf{Bottom:} Representation of the detector and visualisation of an incident muon crossing the detector. }
    \label{fig:sim_tum}
\end{figure}

The main idea  of a detailed tumulus simulation is to compare the real data - atmospheric muons crossing the tumulus - with simulated data using an uniform density tumulus and with no internal structure or monument. Comparing the number of events observed in data and simulation, after a proper normalisation, should reveal in the tumulus any internal structure with a different density than the uniform one, used in the simulation.  A complete simulation of muons crossing the tumulus and detected in the detector requires the following steps :
\begin{itemize}
    \item {The energy and angular distributions of the incoming muons.} 
    
    The two dimensional spectrum, energy versus direction taking into account their correlations, is obtained using a simulation package namely CORSIKA \cite{CORSIKA}. This package tracks incident protons to the top of the atmosphere and follows the generated cascade and therefore muons to the ground. It is known to fit very well all existing data and is largely used by many cosmic ray experiments. However, precise measurements of the muon energy spectra on earth's surface and close to the horizon do not exist and this is one source of systematic errors. The muon energy versus the zenith angle is shown in Figure \ref{fig:EvsTheta}, the zenith angle $ \theta_{z}$ being defined with respect to the vertical and $0^\circ$ representing down-going muons and $90^\circ$ horizontal muons. In first approximation, the zenith angular distribution follows a $\cos^2\theta_{z}$ differential distribution and the azimuthal distribution is isotropic. The CRY cosmic generator \cite{CRY} was also tested, as well as the Reina parametrisation \cite{Reina}    
    \begin{figure}
    \includegraphics[width=15cm]{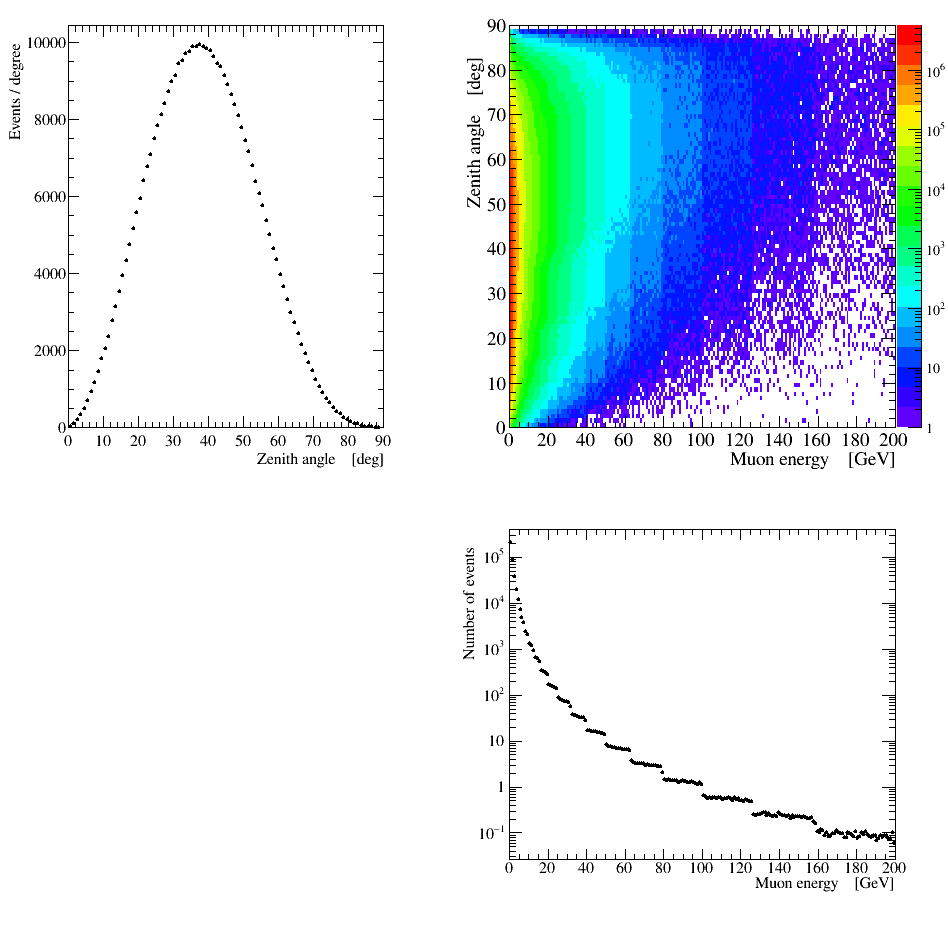}
    \caption{Muon energy versus the zenith angle as generated by CORSIKA. The color code scales with the number of events: the red color materialises bins with the largest number of events.}
    \label{fig:EvsTheta}
    \end{figure}
    
    \item {Muon tracking through the tumulus and detection.}
    
    Generated muons are propagated through the tumulus matter using the GEANT4 package. The tumulus geometry is introduced using precise geodesic measurements performed by geometers and takes into account the real shape of the tumulus. Typically, the tumulus base is 100m wide and the heigtht is 20m. The muon interactions with matter are very well known and simulated. For the tumulus soil composition, a uniform composition of 1\% C, 29\% Si, 15\%Al, 5\%Fe and 50\% O  with a density of $\mathrm{2.2\  g/cm^3}$ is used. Figure \ref{fig:sim_tum}-top displays a muon interacting with matter while crossing the tumulus. 
    \item {Detector simulation.}
    
    The experimental apparatus is completely simulated, starting from its exact geometry, its material and the detector response to an incident muon. For this paper a perfect detector, with no inefficiencies,  is assumed in the simulation. Figure \ref{fig:sim_tum}-bottom shows a muon going through the detector. 
    
\end{itemize}

\section{Experimental setup\label{sec:data}}
A copy of the detector shown in Figure \ref{fig:DirectProblem}-TopLeft was deployed on an experimental site pointing toward a tumulus.
Considering that the height of a tumulus, about 20 m high, the experimental context is less favorable than in the case pyramids presented in section \ref{sec:muography}. Though, the application will serve as a full-scale exercise to deploy high-energy-physics methods in a muography experiment analysis.

\begin{figure}
    \centering
    \includegraphics[trim={0 -15cm 0 0},clip ,width=7.5cm]{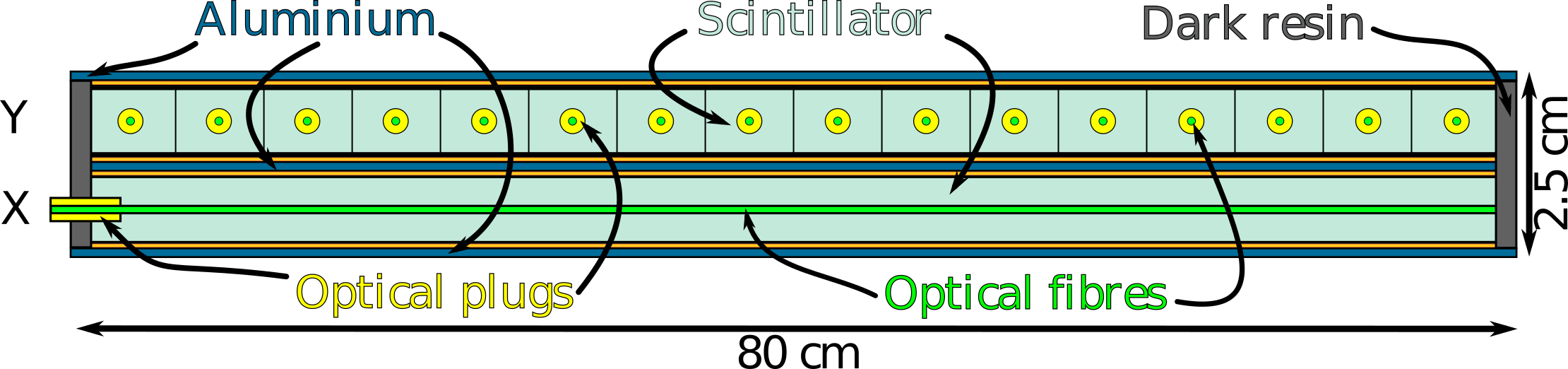} 
    \includegraphics[width=7cm]{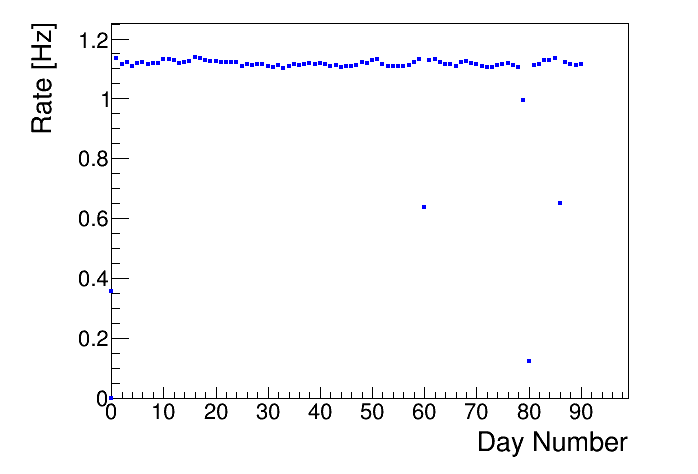}
    \caption{\textbf{Left:} Drawing of a double plane station illustrating the stereo reading. The central fiber is also represented. \textbf{Right:} Monitoring of the data acquisition rate over three months.
    \label{fig:data_det}}
\end{figure}
\subsection{Description of the detector}
The detector is composed of 3 stations spaced 60 cm apart, themselves made up by a double layer of scintillator bars. As shown on the drawing in Figure \ref{fig:data_det}-left, the scintillator bars are aligned in orthogonal directions to form a stereo detection matrix. Each layer is constituted of 32 (resp. 16 ) 80 cm long bars of plastic scintillator with a rectangular section of  $0.5 cm \times 2.5 cm $ (resp. 5cm) for the top and bottom (resp. the central) station. It results in a  32x32 or 16×16 detection matrix. The transverse size of the bars defines the spatial and angular resolution of the detection system and the length of the detector, its geometrical acceptance: a muon going through the detector produces scintillating light in up to six planes.   
The scintillator bars are extruded with a central hole to host a wave-lenght shifter (WLS) fiber for the scintillation light collection.
The WLS optical fibers used, (Kuraray Y11 or Bicron BCF 91A) have $\sim$ 1 mm diameter, collect the UV scintillation light and reemit the signal in the green range where the photosensors have the optimal response. 
A fiber-to-pixel connection in ensured by an optical system which is plugged, a pixelized photo-multiplier (PMT). Those PMTs (Hamamatsu 8804-300)  have $8\times8$ pixels, a typical gain of 106 with a factor 1:3 dispersion on the pixels gains. The relatively high gain requires moderate amplification but the spread implies the necessity of a channel-to-channel gain correction, included in the present electronics of the detector. Each pixel with a sizeable signal is called a hit.  
The coordinate system attached to the detector is represented in Figure \ref{fig:DirectProblem}-TopLeft: $\phi_{Det}$ is measured in X-Y plane where $0^\circ$ is aligned with the X axis. $\theta_{Det}$ is measured in the X-Z plane where $0^\circ$ is aligned with the X axis.

\subsection{Data taking and data preparation}

An initial data acquisition phase took place before the tomography data taking. The detector is oriented vertically so that the detector layers are horizontal: this position maximizes the acceptance to atmospheric muons. Data were acquired for 21.5 hours. This data set is used as calibration data  to check the performance of the setup, such as the inefficiency due to the thresholds used at the level of the PMTs electronics. 
The detector acquisition was set in a two fold coincidence between stations with a coincidence time window of 200 ns where a signal in a station corresponds to at least a double hit in X and Y planes. 

After this initial phase, the detector is rotated towards the tumulus with an inclination of $10^{\circ}$ with respect to the ground. 
The entire data acquisition period lasted for $\sim90$ days of uninterrupted data taking. The detector was monitored daily on site and remotely: Figure \ref{fig:data_det}-right shows the stability of the system over the period. 

Data are pre-processed taking into account only the 3-fold coincidences, and a minimum signal is required in the top and bottom station.


\subsection{Reconstruction of tracks} 
\begin{figure}
    \includegraphics[width=7cm]{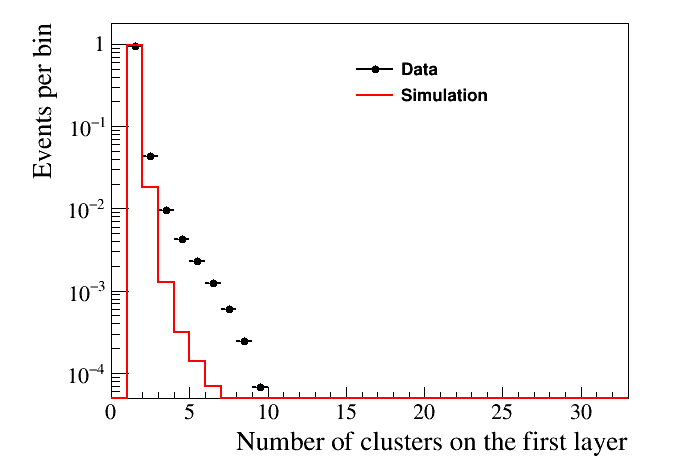}
    \includegraphics[width=7cm]{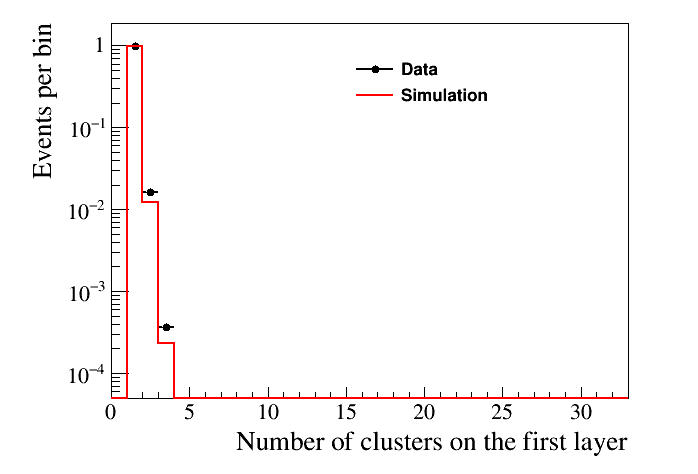}
    \caption{Distribution of the number of clusters on the first layer \textbf{Left:} before and \textbf{Right:} after electron suppression.
    \label{fig:rec_clustermult}}
\end{figure}

Following the data preselection described above, some additional quality cuts are applied to remove events on the edge of the detector. Then neighbouring hits are clustered on each layer. Figure \ref{fig:rec_clustermult}-left shows the distribution of the number of clusters on the first layer in the data and compared with the simulation. 
The significant larger multiplicity observed in data are due to electrons present in cosmic-rays: thanks to the simulation of electrons using GEANT4, their effect in the detector could be reproduced and studied. Their contribution was suppressed by a cut on the total energy deposited in the detector. The result of this cut can be observed in Figure \ref{fig:rec_clustermult}-right: data and simulated muon data are in agreement. 
Next, events with at least one cluster on each layer are kept: using the maximum energy hit as the position of the cluster, a straight line is adjusted in each view providing the azimuthal and zenith angles, $\phi_{Det}$ and $\theta_{Det}$. If there are more than one track in a view, the track with the best $\chi^2$ is kept and only events with one good track in each projection are selected. 
This selection results in around 4.2 millions tracks.

\begin{figure}
    \includegraphics[width=7cm]{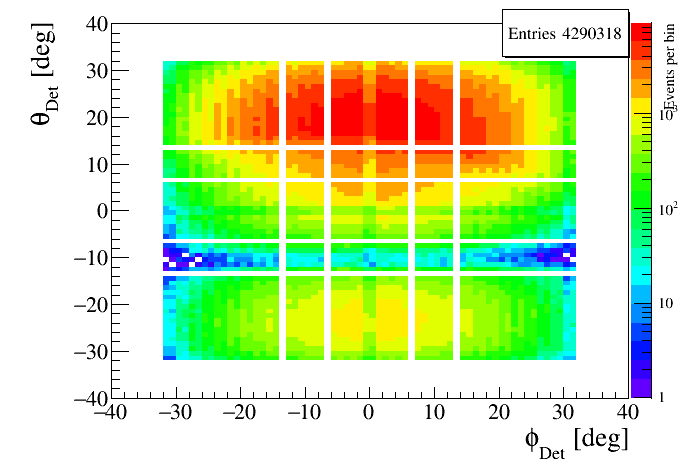}
    \includegraphics[width=7cm]{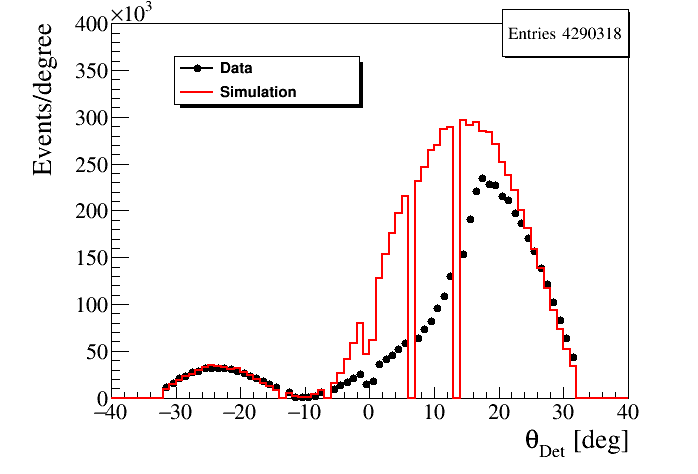}
    \caption{\textbf{Left:} Data distribution of $\theta_{Det}$ versus $\phi_{Det}$ angles of reconstructed tracks. \textbf{Right:} Distribution of $\theta_{Det}$ in data and simulation; the uneven distribution is explained by the segmentation of the detector. \label{fig:rec_events}}
    
\end{figure}

Figure \ref{fig:rec_events} represents the distribution in $\theta_{Det}$ and $\phi_{Det}$ measured in the detector coordinates. 
On the projected $\theta_{Det}$ distribution, three features can be identified :
\begin{itemize}
    \item The inclination of the detector is visible at $\theta_{Det}\ \approx\ -10^\circ$.
    \item Tracks with $\theta_{Det}\ <\ -10^{\circ}$ represent backward tracks entering the detector by the last station. These tracks can be used to normalise the simulation. 
    \item Finally, the deficit of tracks around $\theta_{Det}$ equals $+10^{\circ}$ to $+20^{\circ}$ highlights the absorption of muon tracks in the tumulus, with respect  to open sky. 
\end{itemize}

\section{Data and Simulation comparison \label{sec:results}} 
Using the angular and energy distributions, obtained with CORSIKA, a very large number of muon events reaching the detector volume was generated; low energy muons are absorbed by the tumulus as a  muon loses in average 600 MeV per meter of soil crossed. Typically, horizontal muons with energy less than 60 GeV are stopped in the tumulus. In addition incident muons will be scattered by the tumulus and will change direction. 
\par
In order to allow a direct comparison between data and simulated data, it is important to get the right normalisation between the two samples. 
\subsection{Normalisation using open-sky data}
\begin{figure}
   \hspace*{-1.2cm}
    \includegraphics[width=9.5cm]{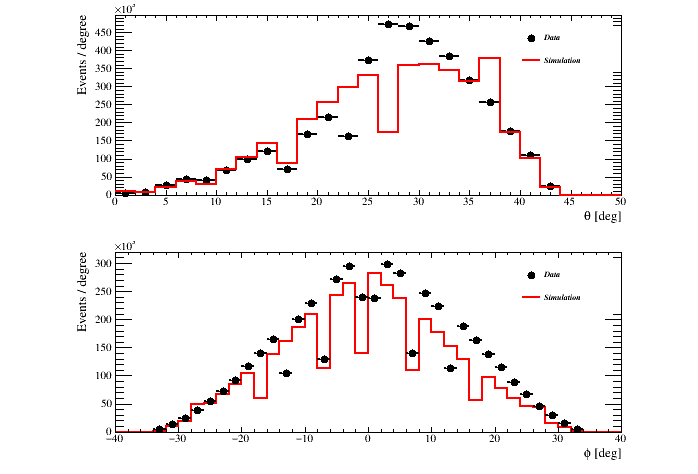}
    \hspace*{-1.cm}
    \includegraphics[width=9.5cm]{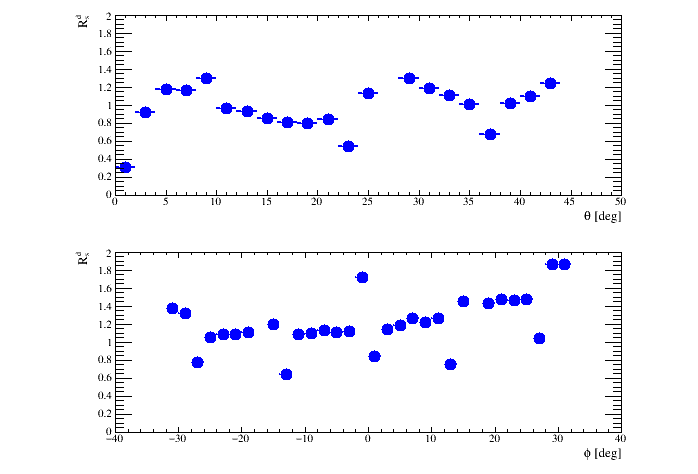}
    \caption{\textbf{Top left:} $\theta$ distribution of events observed in data and simulation. \textbf{Bottom left:} $\phi$ distribution of events in data and simulation.
    \textbf{Top right:} Ratio $R^{data}_{sim}$ versus $\theta$. \textbf{Bottom right:} Ratio $R^{data}_{sim}$ versus $\phi$
        \label{fig:DMC1D}}
\end{figure}

\begin{figure}
    \includegraphics[width=14cm]{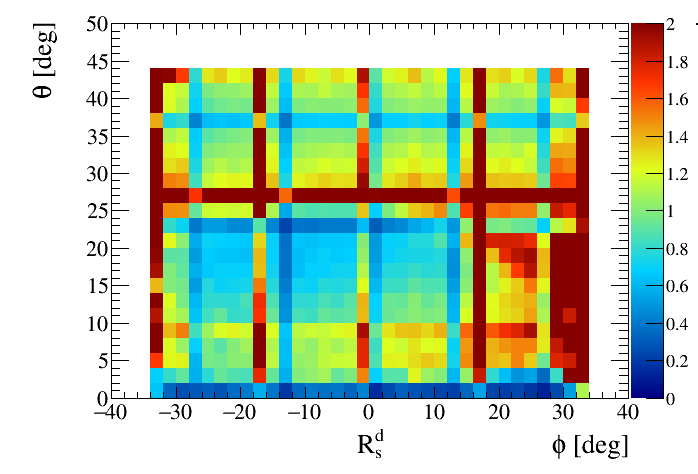}
    \caption{Ratio $R^{data}_{sim}$ versus azimuthal and polar angles }
        \label{fig:DMC2D}
\end{figure}

A large number of muons that reach the detector do not cross the tumulus and come from the open sky. This is the case for those with a polar angle $\theta > 32^{\circ} $.
For practical reasons, the polar angle $\theta$ used in this section is defined as $\theta_{Det}\ +\ 10^{\circ}$, and the azimuthal angle, $\phi$ is equal to $\phi_{Det}$.   
After checking that the angular distribution above  $32^{\circ} $ is well reproduced by the simulation, the normalisation factor is obtained from the ratio between the number of observed events above $32^{\circ} $ divided by the number of simulated events in the same angular range.
Figure \ref{fig:DMC1D}-left shows the $\theta$ and $\phi$ distributions of events in data and in simulation after normalising the simulation with the factor defined above. In the open sky region, $\theta > 32^{\circ}$, a good agreement is observed. For $\theta \sim 25^{\circ} $, more events are observed in the data sample showing probably that the tumulus height in the simulation is higher than in reality, absorbing more muons. For $\theta < 20^{\circ} $, and tracks crossing in majority the tumulus the agreement is fair. The ratio $R^{data}_{sim}$ is shown in Figure \ref{fig:DMC1D}-right: the agreement between data and simulation is not better than $20\%$ in average.  
Figure \ref{fig:DMC2D} shows the ratio $R^{data}_{sim}$ in two dimensions, polar angle versus azimuthal angle, in bins of $2^{\circ}$ allowing a fine image of the tumulus. Any hidden monument should absorb more muons resulting in a ratio $R^{data}_{sim}<1$ at its position inside the tumulus. For $\phi>25^{\circ} $and $\theta < 20^{\circ}$, much more data than simulated data are observed, meaning that the tumulus shape or orientation in the simulation lack a precise description.The precision of the simulation, as of today, does not allow to draw any relevant conclusions. The other two cosmic generator or parametrisation mentioned in section \ref{sec:mc} exhibit a more severe disagreement with data than CORSIKA. 

\subsection{Normalisation with the backward tracks} 
Another possibility, expected to be less dependant on the precision of the simulation is to normalize the distribution using the backward tracks: these tracks are in an angular range which is similar to those crossing the tumulus contrary to open-sky tracks. 
To illustrate this alternative method, events were simulated isotropically in $\phi$ and following a pure $cos^2$ differential distribution in $\theta$, the energy correlation with $\theta$ being not taken into account; the simulated events were generated on 2 virtual planes placed symmetrically in front and behind the detector: these tracks do not cross the tumulus. 
The red line in Figure \ref{fig:rec_events}-right shows the simulated distribution normalized to the number of backward tracks. It can be seen that the open sky distribution is reasonably well reproduced ($\theta_{Det} > 20 \deg$): this observation reinforces the confidence in the more precise CORSIKA simulation used above. On the same figure, the deficit of tracks due to the absorption in the tumulus is clear. To use this normalisation it is necessary in the future to generate backward tracks at the same time as forward tracks, using the full GEANT4 simulation.  

\section{Conclusions and outlook}
A proof of concept experiment was performed taking cosmic muons data with a 3-station muon detector pointing towards a tumulus. A full simulation package based on GEANT4, toolkit for high-energy physics, was developed in order to compare with the real data and to reveal internal structures of the tumulus. At this stage, the overall agreement between data and simulation is not better than $20\%$, insufficient for discovering a monument made by marble whom density $\mathrm{2.5\  g/cm^3}$ is very close to the surrounding soil. The current simulation is limited by the lack of knowledge of the real density of the tumulus soil, of its precise geometry and of the exact detector position relative to the tumulus. The three items will be addressed in the future. In order to improve the precision, more data are needed, meaning a very long data taking time, given the low number of muons at the horizon. The use of two or three 4-layers detectors will enhance the discovery potential, allowing in addition for an online monitoring of the detector. Combining these data with those from other techniques used in geosciences is mandatory to confirm any discovery, and allow archaeologists to excavate the right position.   

\printbibliography

@article{Alvarez,
author={L. W Alvarez et al.},
title={Search for Hidden Chambers in the Pyramids},
journal={Science, New Serie},
volume={167},
year={1970}
}

@article{Scan,
author={K. Morishima et al.},
title={Discovery of a big void in Khufu's Pyramid by observation of cosmic-ray muons.},
journal={Nature},
volume={24647},
year={2015}
}

@article{GEANT1,
author = "The GEANT4 Collaboration",
title = "{GEANT4 -- a simulation toolkit}",
journal = "NIM A",
volume = "506",
year = "2003"}

@unpublished{CORSIKA,
author = "{D. Heck, J. Knapp, J.N. Capdevielle, G. Schatz, T. Thouw}",
title = "CORSIKA",
report-number = "FZKA 6019",
year = "1998"}

@article{CRY,
author = "{C. Hagmann, D. Lange and D. Wright}", 
title = "Cosmic-ray shower generator (CRY) for Monte Carlo transport codes",
journal = "IEEE Nuclear Science Symposium Conference Record, pp 1143-1146",
year = "2007"
%doi = "10.1109/NSSMIC.2007.4437209"
}

@article{Reina,
author="D. Reyna",
title= "A Simple Parametrization pf the Cosmic-Ray Muon Momentum Spectra at the Surface as a function of the Zenith Angle.",
journal = "arXiv:hep-ph/0604145v2",
year= "2006"}

@article{Sullivan,
author = "J. D. Sullivan",
title = "Geometrical factor and directional response of single and multi-element particle telescopes",
journal = "NIM",
volume = "95",
year = "1971"}
\end{document}